\documentclass{webofc}
\usepackage[varg]{txfonts} 

\begin{document}
\title{On the stratifications of 2-qubits X-state space}

\author{\firstname{Arsen} \lastname{Khvedelidze}\inst{1,2,3,4}\fnsep\thanks{\email{akhved@jinr.ru}} \and
        \firstname{Astghik} \lastname{Torosyan}\inst{4}\fnsep\thanks{\email{astghik@jinr.ru}} 
}

\institute{A. Razmadze Mathematical Institute, 
           Iv.Javakhishvili Tbilisi State University,                Tbilisi, Georgia
\and
           Institute of Quantum Physics and Engineering              Technologies, 
           Georgian Technical University, Tbilisi, Georgia
\and
           National Research Nuclear University,
           MEPhI (Moscow Engineering Physics Institute),
           Moscow, Russia
\and
           Laboratory of Information Technologies,  
           Joint Institute for Nuclear Research,
           Dubna, Russia
          }

\abstract{
The 7-dimensional family $\mathfrak{P}_X$ of so-called mixed X-states of 2-qubits is considered.  
Two types of stratification of 2-qubits $X-$state space,  i.e., partitions   of  $\mathfrak{P}_X\,$ into orbit types with respect to the adjoint group actions, one of the global unitary group $G_X \subset SU(4)$ and another one under the action of the local unitary group $LG_X \subset G_X$,  is described.   
The equations and inequalities in the invariants of the corresponding groups, determining  each stratification component, are given.
}

\maketitle

\section{Introduction}
\label{intro}
The understanding of a symmetry that a physical system possesses, as well as this symmetry's breaking pattern allows us to explain uniquely a wide variety of phenomena in many areas of physics, including elementary particle physics and condensed matter physics \cite{Michel}. The mathematical formulation of symmetries related to the Lie group action consists of the detection of the stratification of the representation space of corresponding symmetry group. Dealing with closed quantum systems the symmetries are realized by the unitary group actions and the quantum state space plays the role of the symmetry group representation space.  Below, having in mind these observations,  we will outline examples of the stratifications occurring for a quantum system composed of a pair of 2-level systems, two qubits. We will analyze symmetries associated with two subgroups of the special unitary group $SU(4)$. More precisely,  we will consider the
7-dimensional subspace $\mathfrak{P}_X$ of a generic 2-qubit state space, the family of $X$-states (for definition see \cite{YuEberly2007}, \cite{AAJMS2017} and references therein) and reveal two types of its partition into the set of points having the same symmetry type. The  primary stratification originates from the action of the invariance group of $X$-states, named the global unitary group $G_{{}_X} \subset SU(4)$,  whereas  the secondary one is  due to the action of the so-called local group $LG_{{}_X}\subset  G_X$ of the $X$-states. 

%%%%%%%%%%%%%%%%%%%%%%
\section{X-states and their symmetries}
\label{sec-1}
%%%%%%%%%%%
The mixed 2-qubit  
$X\--$states can be defined based on the purely algebraic consideration. The idea is to fix the  subalgebra 
$
\mathfrak{g}_X:=\mathfrak{su}(2)\oplus\mathfrak{su}(2) \oplus\mathfrak{u}(1) \in \mathfrak{su}(4)
$ 
of the algebra $\mathfrak{su}(4)$ 
and define the density matrix of $X-$states as 
\begin{equation}\label{eq:XmatrixAlg}
\varrho_X= \frac{1}{4}\left(I + i\mathfrak{g}_X \right)\,.
\end{equation}
In order to coordinatize the $X\--$state space we use the  tensorial basis for the 
$\mathfrak{su}(4)$ algebra, $\sigma_{\mu\nu} =\sigma_\mu\otimes\sigma_\nu, \ \mu,\nu =0,1,2,3$. It consists of all possible tensor products of two copies of Pauli matrices and a unit $2\times 2$ matrix, $\sigma_\mu=(I, \sigma_1,\sigma_2,\sigma_3)\,,$ which we order as follows (see the details in 
\cite{AAJMS2017}):
\begin{eqnarray}
&&\lambda_1, \dots, \lambda_{15} = 
\frac{i}{2}\left(\sigma_{x0},
\sigma_{y0}, \sigma_{z0}, \sigma_{0x}, \sigma_{0y}, \sigma_{0z},\sigma_{xx},
\sigma_{xy}, \sigma_{xz}, \sigma_{yx}, \sigma_{yy}, \sigma_{yz},
\sigma_{zx}, \sigma_{zy}, \sigma_{zz} \right)\,.
\end{eqnarray}
In this basis the 7-dimensional subalgebra 
$\mathfrak{g}_X $ is generated by the subset 
$\alpha_X=\left( \lambda_{3}, \lambda_{6}, \lambda_{7},  \lambda_{8}, \lambda_{10}, -\lambda_{11}, \lambda_{15} \right),$ and thus the unit norm $X\--$state density matrix is given by the decomposition:
\begin{equation}\label{eq:XmatrixExp}
\varrho_X= \frac{1}{4}\left(I +2 i\sum_{\lambda_k \in \alpha_X} h_k\lambda_k\right)\,.
\end{equation}
The real coefficients $h_k$ are subject to the   polynomial inequalities ensuring the semi-positivity of the density matrix, $\varrho_X \geq 0:$
\begin{equation}\label{eq:positivity}
\mathfrak{P}_X= \{h_i \in \mathbb{R}^7 \ | \ \left(h_3\pm h_6\right){}^2+\left(h_8\pm h_{10}\right){}^2+\left(h_7\pm h_{11}\right){}^2\leq (1\pm h_{15})^2 \}\,. 
\end{equation}
Using the definition (\ref{eq:XmatrixAlg}) one can conclude that the $X-$state space  $\mathfrak{P}_X$ is invariant under the 7-parameter group, 
$ G_X :=\exp (\mathfrak{g}_X)\in SU(4):$
\begin{equation}
g\varrho_X g^\dag \in  \mathfrak{P}_X\,, \qquad \forall g \in G_X\,.
\end{equation}
Group $G_X$ plays the same role for the $X\--$states as  the special unitary group $SU(4)$ plays for a generic 4-level quantum system, and thus is termed the \textit{global unitary group} of  $X\--$states. According to \cite{AAJMS2017},  group $G_X$ admits the representation:
\begin{equation}\label{eq:G_XRepr}
G_X=P_{\pi}\left(
\begin{array}{c|c}
{e^{-i {\omega_{15}}}SU(2)}& 0  \\
\hline
0 &{e^{i {\omega_{15}}}SU(2)^\prime}\\
\end{array}
\right)P_{\pi}\,,\quad \mbox{with}\quad 
P_{\pi}=\left(\begin{matrix}
1&0&0& 0\\
0&0&0&1\\
0&0&1&0\\
0&1&0&0
\end{matrix}
\right)\,.
\end{equation}
Correspondingly,  the \textit{local unitary group} of the $X\--$states is 
\begin{equation}\label{eq:LG_XRepr}
LG_X=P_{\pi}\exp(\imath \frac{\phi_1}{2})
\otimes
\exp(\imath \frac{\phi_2}{2})P_{\pi}\ \subset G_X\,.
\end{equation}
%%%%%%%%%%%%%%%%%%%%%%%%%%%%%

%%%%%%%%%%%%%%%%%%%%%%%%%%%%%%%%%%%%
\section{Global orbits and state space decomposition}
\label{sec-2}

\begin{figure}
\centering
\sidecaption
\includegraphics[width=5cm,clip]{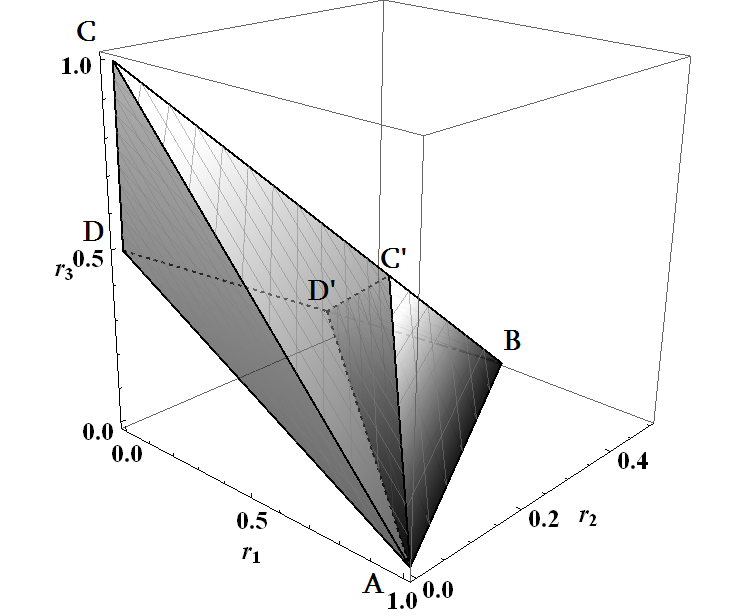}
\caption{The tetrahedron $ABCD$ describes the partially ordered simplex  $\underline{\Delta}_3:= \{ \  \sum_{i=1}^4 r_i =1\,, \ \{1 \geq r_1 \geq r_2 \geq 0 \}\cup\{ 
1 \geq r_3 \geq r_4 \geq  0\} \}$ of the density matrix eigenvalues,  while the tetrahedron $ABC'D'$ inside  it  corresponds to a 3D simplex with the following complete order: 
 $ \{  \  \sum_{i=1}^4 r_i =1 \,, \;   1 \geq r_1 \geq r_2 \geq r_3 \geq r_4 \geq  0\ \} \,$.}{ \label{Fig:PartOrderedSimplex}}
\label{fig-3}    
\end{figure}

Now we give a classification of the global $G_X$-orbits according to their dimensionality and isotropy group.  
Every density matrix $\varrho_X $ can be diagonalized by some element of the global $G_X$ group. In other words, all global $G_X$-orbits can be generated from the  density 
matrices, whose eigenvalues form the partially ordered simplex  $\underline{\Delta}_3$, depicted on Figure \ref{Fig:PartOrderedSimplex}. 

The tangent space to the $G_X$-orbits is spanned by the subset of linearly independent vectors, built from the  vectors: 
\(
t_k = [\lambda_k,\varrho_X], \ \lambda_k \in \alpha_X\,.
\)
The number of independent vectors $t_k$ determines the dimensionality  of the $G_{{}_X}$-orbits and is given  by the rank of the  $7 \times 7$ Gram matrix:
\begin{equation}\label{eq:Gram}
\mathcal{G}_{kl} = \frac{1}{2} \mbox{Tr}(t_k t_l)\,.
\end{equation}
The Gram matrix (\ref{eq:Gram}) has three zero eigenvalues and two double multiplicity eigenvalues:
\begin{eqnarray}
\mu_\pm&=& -\frac{1}{8}\left(\left(h_3\pm h_6\right){}^2+\left(h_8\pm h_{10}\right){}^2+\left(h_7\pm h_{11}\right){}^2\right)\,.
\end{eqnarray}
Correspondingly, the $G_X$-orbits  have  dimensionality of  either 4, 2 or 0.
The orbits of maximal dimensionality, {$\mbox{dim}\left( {\mathcal{O}}\right)_{\mbox{\small Gen}}=4$}, are characterized  by non-vanishing 
$\mu_\pm\neq 0$ and consist of the set of density matrices with a generic spectrum, $\sigma(\varrho_x)= (r_1,r_2,r_3,r_4)\,.$
If the density matrices  obey the equations  
\begin{equation}
h_6=\pm h_3\,, \  h_{10}=\pm h_8\,, \  \  h_{11}=\pm h_7\,,
\end{equation}
they belong to the so-called degenerate orbits, 
$\mbox{dim}\left({\mathcal{O}}\right)_{\pm}=2\,.
$
The latter are generated from the matrices which have the double degenerate spectrum of the  form, $\sigma(\varrho_x)= (p,p,r_3,r_4)$ and  $\sigma(\varrho_x)= (r_1,r_2,q,q)$ respectively. 
Finally, there is a single orbit 
$\mbox{dim}\left({\mathcal{O}}\right)_{0}= 0$, corresponding to the maximally mixed state   $\varrho_X = \frac{1}{4}I$. 

Considering the diagonal representative of the generic $G_{{}_X}$-orbit one can be convinced that its  \textit{isotropy group} is  
 \begin{equation}
 H=P_{\pi}\left(
\begin{array}{c|c}
e^{i\omega}\exp{i{\frac{\gamma_1}{2}\sigma_3}}& {}^{\mbox{\Large  0 }} 
\\\hline
 {}_{\mbox{\Large  0 }}&e^{-i\omega}\exp{ i{\frac{\gamma_2}{2}\sigma_3}}\\
\end{array}
\right) P_{\pi}\,, 
 \end{equation}
 while for a diagonal representative
 with a double degenerate spectrum the isotropy group is given by one of two groups:
 \begin{equation*}
 H_{+}=P_{\pi}\left(
\begin{array}{c|c}
e^{i\omega}SU(2)& {}^{\mbox{\Large  0 }} 
\\\hline
 {}_{\mbox{\Large  0 }}&e^{-i\omega}\exp{i{\frac{\gamma_2}{2}\sigma_3}}\\
\end{array}
\right) P_{\pi}\,, \quad
 H_{-}=P_{\pi}\left(
\begin{array}{c|c}
e^{i\omega}\exp{i{\frac{\gamma_1}{2}\sigma_3}}& {}^{\mbox{\Large  0 }} 
\\\hline
 {}_{\mbox{\Large  0 }}&e^{-i\omega}SU(2)^\prime\\
\end{array}
\right) P_{\pi}\,. 
 \end{equation*}
 For the single, zero dimensional orbit the isotropy group $H_0$ coincides with the whole invariance group, $H_0=G_X$. 
Therefore, the isotropy group of any element of 
$G_{{}_X}$-orbits belongs to one of these conjugacy classes: $[H], [H_\pm]$ or $[H_{0}]$. Moreover, a straightforward analysis shows that 
$[H_{+}]=[H_{-}].$
Hence, any point $\varrho \in  \mathfrak{P}_X$ belongs to one of three above-mentioned types of $G_{{}_X}$-orbits\footnote{
The \textit{orbit type} $[\varrho]$ of a point $\varrho \in \mathfrak{P}_X$ is given by the conjugacy class of the isotropy group of point $\varrho,$ i.e.,  $[\varrho]=
[G_{\varrho_X}] 
$\,.
}, denoted afterwards as $[H_{t}]\,, t =1,2,3.$ 
For a given $H_t$, the associated \textit{stratum}   $\mathfrak{P}_{[H_t]}\,,$ defined as 
the set of all points whose stabilizer is conjugate to $H_t$: 
$$\mathfrak{P}_{[H_t]}: =\{
y \in \ \mathfrak{P}_X|\  
\mbox{isotropy~group~of~} y \mbox{~is~conjugate~to}\  H_t\}
$$
determines the  sought-for decomposition 
of  the state space $\mathfrak{P}_{X}$ into strata according to the  orbit types:
\begin{equation}
{\mathfrak{P}}_X =\bigcup_{\mbox{orbit types}}{\mathfrak{P}}_{[H_i]}\,.
\end{equation}
The  strata  ${\mathfrak{P}}_{(H_i)}$  are determined by this set of equations and inequalities: 
\begin{eqnarray}
&(1)& {\mathfrak{P}}_{[H]}:  = 
\{h_i \in \mathfrak{P}_X \ |\  \mu_+ > 0,\, \mu_- > 0\,\}\,,\\
&(2)& {\mathfrak{P}}_{[H_+]}\cup {\mathfrak{P}}_{[H_-]} := \{h_i \in \mathfrak{P}_X \ |\  \mu_+ = 0,\, \mu_- > 0\,\}\cup \{h_i \in \mathfrak{P}_X \ | \ \mu_+  > 0,\, \mu_- = 0\,\}\,,\\
&(3)& {\mathfrak{P}}_{[H_0]}: = \{h_i \in \mathfrak{P}_X \ |\  \mu_+ = 0,\,  \mu_- = 0\,\}\,.
\end{eqnarray}
%%%%%%%%%%%%%%%%%%%

%%%%------------------------------
\section{Local orbits and state space decomposition}
%%%%%%%%%%%%%%-------------------------

Analogously,  one can build up the $X\--$ state space decomposition associated with the local group $LG_{{}_X}$ action. For this action the dimensionality of $LG_{{}_X}$-orbits is given by the rank of the corresponding  $2 \times 2$ Gram matrix constructed out of vectors $t_3$ and $t_6$. Since its eigenvalues read:
\begin{eqnarray}
\mu_1= -\frac{1}{8}\left(\left(h_8+h_{10}\right){}^2+\left(h_7+h_{11}\right){}^2\right)\,,\qquad
\mu_2=-\frac{1}{8}\left(\left(h_8-h_{10}\right){}^2+\left(h_7-h_{11}\right){}^2\right)\,,
\end{eqnarray}
the $LG_X$-orbits are either generic ones with the dimensionality of {$\mbox{dim}\left( {\mathcal{O}_L}\right)_{\mbox{\small Gen}}=2$}, or degenerate {$\mbox{dim}\left({\mathcal{O}_L}\right)_{\pm}=1$}, or exceptional ones, {$\mbox{dim}\left({\mathcal{O}_L}\right)_{0}=0$}.
The $LG_X$-orbits can be collected into the strata according to their orbit type. There are three types of strata associated with the ``local'' isotropy subgroups of $LH \in LG_X \,$. Correspondingly, one can define the following ``local'' strata of state space: 
\begin{itemize}
\item the generic stratum, ${\mathfrak{P}}^L_{[I]}$, which  has  a trivial isotropy type, $[I],$  and is represented by  the inequalities: 
${\mathfrak{P}}^L_{[I]}:  = 
\{h_i \in \mathfrak{P}_X \ |\  \mu_1 > 0,\,  \mu_2 > 0\,\}\,,
$
\item the degenerate stratum, ${\mathfrak{P}}^L_{[H_L^\pm]}$, collection of the  orbits whose type is $[H_L^\pm]\,,$ with the subgroup either   $H_L^+= I\times\exp{(iu\sigma_3)},$ or  $H_L^-= \exp{(iv\sigma_3)}\times I$. The stratum defining equations read respectively:
\begin{equation}
h_{10} = \pm h_8\,, \ \quad  \  h_{11} = \pm h_7\,,
\end{equation}
\item the exceptional stratum, ${\mathfrak{P}}_{[LG_X]}$  of the type $[LG_X]\,$, determined by the equations:  $h_{11}=h_{10}=h_8=h_7=0\,$. 
\end{itemize}
Therefore, the local group  action  prescribes the following stratification of 2-qubit $X\--$state space: 
\begin{equation}
{\mathfrak{P}}_X ={\mathfrak{P}}_{[I]}\cup{\mathfrak{P}}_{[H^+_L]}\cup{\mathfrak{P}}_{[H^-_L]}\cup{\mathfrak{P}}_{[LG_X]}\,.
\end{equation}

%%%%%%%%%%%%%%%%%%%%%%%%
\section{Concluding remarks}
%%%%%%%%%%%%%%%%%%%%%%%%

In the present article we describe the stratification of 2-qubit $X\--$state space associated with the adjoint action of the global and local unitary groups. The global unitary symmetry is related to the properties  of a system as a whole, while the local symmetries comprise information on the entanglement, cf. \cite{Chen2010}. In an upcoming publication, based on the introduced stratification of state space, we plan to analyze an interplay between these two symmetries and particularly determine the entanglement/separability characteristics  of every stratum. 

%%%%%%%%%%%%%%%%

\end{document}